# Mutual Spin-Phonon Driving Effects and Phonon Eigenvector Renormalization in Nickel (II) Oxide.

Qiyang Sun[1], Bin Wei[1,3], Yaokun Su[2], Hillary Smith[4], Jiao Y. Y. Lin[5], Douglas L. Abernathy[6], and Chen Li[1,2]*

[1]Department of Mechanical Engineering, University of California, Riverside, Riverside, CA 92521, USA

[2]Materials Science and Engineering, University of California, Riverside, Riverside, CA 92521, USA

[3]Henan Key Laboratory of Materials on Deep-Earth Engineering, School of Materials Science and Engineering, Henan Polytechnic University, Jiaozuo 454000, China

[4]Department of Physics and Astronomy, Swarthmore College, Swarthmore, PA 19081, USA

[5]Neutron Data Analysis and Visualization Division, Oak Ridge National Laboratory, Oak Ridge, TN 37830, USA

[6]Neutron Scattering Division, Oak Ridge National Laboratory, Oak Ridge, TN 37830, USA

**Abstract**

The physics of mutual interaction of phonon quasiparticles with electronic spin degrees of freedom, leading to unusual transport phenomena of spin and heat, has been a subject of continuing interests for decades. Despite its pivotal role in transport processes, the effect of spin-phonon coupling on the phonon system, especially acoustic phonon properties, has so far been elusive. By means of inelastic neutron scattering and first-principles calculations, anomalous scattering spectral intensity from acoustic phonons was identified in the exemplary collinear antiferromagnetic nickel (II) oxide, unveiling strong spin-lattice correlations that renormalize the polarization of acoustic phonon. In particular, a clear magnetic scattering signature of the measured neutron scattering intensity from acoustic phonons is demonstrated by its momentum transfer and temperature dependences. The anomalous scattering intensity is successfully modeled with a modified magneto-vibrational scattering cross section, suggesting the presence of spin precession driven by phonon. The renormalization of phonon eigenvector is indicated by the observed "geometry-forbidden" neutron scattering intensity from transverse acoustic phonon. Importantly, the eigenvector renormalization cannot be explained by magnetostriction but instead, it could result from the coupling between phonon and local magnetization of ions.

**Significance Statement**

Nickel (II) oxide is a prominent candidate of antiferromagnetic spintronic applications, largely thanks to its high Néel temperature. We present scattering signatures of mutual driving interactions through strong spin-lattice coupling and acoustic phonon eigenvector renormalization in this important material for the first time. Our results provide a new approach to identify and quantify strong spin-phonon interactions, shedding lights on engineering functional spintronic and spin-caloritronic materials through these interactions.

**Main Text**

**Introduction**

Phonons, as the quanta of lattice vibrations, are known to strongly couple with spin and electronic degrees of freedom in a variety of magnetic materials, leading to many intriguing novel phenomena. In magnetic insulators, the large electron band gap accompanied with high energy of crystal electric field excitations prevents the direct coupling of phonon or spin system to the electronic system, whereas the transport processes are affected predominately by spin-phonon coupling. Unusual spin and phonon transport phenomena, including phonon pumping of spin current [1-5] and pumping of phonon by magnetization dynamics [6, 7], suggest phonon propagation can greatly affect the transport of spin and vice versa, and have made spin-phonon dynamics an active field of research. However, some aspects of the interaction between lattice and spin degrees of freedom are still unclear, e.g., the selection rule for spin-phonon interactions, the underlying mechanism of spin dissipation through lattice, quantification of spin-phonon coupling strength. In contrast to optical phonons, acoustic phonons with large group velocity play an important role in transport process. Resolving acoustic phonon characteristics in magnetic insulators will help understand spin-phonon interactions and engineer functional spintronic [8] and spin-caloritronic [9, 10] devices.

Antiferromagnetic (AFM) insulators have been of particular interest for applications in next-generation signal processing devices due to their ultra-low dissipation in spin transportation. The collinear AFM nickel (II) oxide (NiO) with a Néel temperature ($T_N$) of 523 K, is a promising candidate material for spintronic and spin-caloritronic applications at ambient temperature, thanks to its high efficiency in spin transport [11-14]. It has a simple face center cubic (FCC) structure in the paramagnetic (PM) phase and takes on a slight rhombohedral 0.09° distortion (deviating from 60°) in the AFM phase at 5 K [15]. While the AFM spin order together with the asymmetry of Born effective charge on $Ni^{2+}$ ion gives rise to the splitting of two transverse optical (TO) phonons [16, 17] that are degenerate in FCC crystal symmetry, their effects on acoustic phonon remains unexplored. Moreover, the increase of thermal conductivity beyond $T_N$ in NiO [18] hints at the existence of strong spin-phonon interactions and motivates the current work.

The most well-known spin-phonon coupling process is the modulation process, which refers to the dynamic modulation of exchange coupling between magnetic ions induced by phonons. While the spin-phonon coupling is known to result in modifications on magnon dispersions by the modulation process [19, 20], the effect on phonon system is less well understood. For example, the spin-driven renormalization of phonon energy is only phenomenologically characterized by a spin-phonon coupling coefficients in many magnetic systems [21-23]. The interaction of magnons and acoustic phonons has been extensively studied in various magnetic systems [24-29], revealing the formation of hybrid magneto-acoustic modes, which manipulate the acoustic phonon dispersion and pass angular momentum to acoustic phonons [30]. However, existing observations are constrained to the vicinity of crossing points of magnon-phonon dispersions, and do not provide the effects of such interactions on whole dispersion branches. Inelastic neutron scattering (INS) measurements on single crystals have been proved to be a powerful probing technique in resolving acoustic phonon characteristics. Previous INS experiments reveal that phonon INS intensity can be modified with spin-phonon coupling and "magneto-vibrational" modes, which do not follow the wavevector (**Q**) and temperature dependences of phonons, are formed [31-35]. Therefore, a theoretical simulation of phonon dynamic structure factor will shed light on identifying such modes and their scattering origin in a system with strong spin-phonon coupling.

Here, we report INS experiments and atomistic simulations that demonstrate the existences of mutual spin-phonon driving effects and the renormalization of acoustic phonon eigenvectors in bulk NiO. A strong INS intensity that follows acoustic phonon dispersions is observed at low **Q** region, where the lattice INS cross section is small, suggesting its magnetic origin. The unusual temperature dependence of such intensity is associated to the temperature dependence of magnetic order, suggestive of a strong coupling between phonon and spin. The observed intensity at low **Q** is successfully reproduced by magneto-vibrational scattering cross section based on a strong magneto-elastic correlated picture, indicating the presence of spin precession driven by phonon. More strikingly, strong INS intensity from "geometry-forbidden" transverse acoustic (TA) phonons is observed at high **Q** and shown to originate from lattice INS, indicating the renormalizations of acoustic phonon eigenvectors. The renormalization process can be related to the coupling between phonon and local magnetization and suggestive of spin driven effects on lattice vibration.

Time-of-flight INS measurement are performed on both single crystal and polycrystalline NiO (see Methods) to measure the dynamic structure factor $S(\mathbf{Q}, E)$. The simulation of dynamic structure factor (see Supplementary Information) based on first-principles phonon calculations (see Methods) has a good agreement with measured phonon dispersions, as shown in Fig.1 a,b and Fig.S1 a-d in Supplementary Information.

**Results and Discussion**

Anomalously large INS intensity that follows acoustic phonon dispersion is observed at low **Q** and has magnetic origin. In Fig.1 a, pronounced spectral intensity below 45 meV is observed in the first BZ ($0 < L < 1$) along $[0, 0, 1]$ in AFM phase. Such intensity follows the dispersion of acoustic phonons, but it does not follow the $\mathbf{Q}^2$ dependence of coherent one-phonon INS process, as shown by the simulation in Fig.1 b. Instead, strong intensity is shown at $0 < L < 1$, with the mode weakening at $1 < L < 3$ and strengthening again at $L > 3$. Such trend is visualized by constant **Q** cuts at various equivalent reciprocal points, as presented in Fig.1 d,e. This is highly unusual because phonon INS cross section is expected to be smaller at low **Q**. The anomalously strong intensity and its **Q** dependence suggest that it cannot arise solely from lattice scattering. On the other hand, the magnetic INS cross section, which is subject to the $Ni^{2+}$ magnetic form factor, is expected to decrease with **Q**. Therefore, the intensity of theses modes at $L < 4$ can be understood as a combined contribution from magnetic and lattice scattering processes, where magnetic scattering gives diminishing intensity at higher **Q** while lattice scattering increases with **Q**. These observations also suggests that the acoustic-phonon-like intensity in the first BZ ($L < 2$) may predominantly originate from magnetic scattering. For simplicity, such modes will still be referred as phonon modes.

Interestingly, strong spectral intensities from acoustic phonon modes are also observed in the first BZ along $[-1, 1, 1]$ and $[-1, 1, 0]$, as shown in Supplementary Information (Fig.S1, Fig.S2), indicating such behavior was not limited to $[0, 0, 1]$ direction. To elucidate the appearance of anomalous phonon INS intensity in small **Q** regime, volumetric views of simulated (one-phonon) and measured dynamic structure factors at 10 K are shown in Fig.2 a,c. The LA, TO, and LO phonon branches are well captured by the lattice dynamic structure factor calculation. At 10 K in the vicinity of $(0, 0, 2)$ and $(-2, 2, 0)$, the calculated intensity of LA modes is 2 orders of magnitude stronger than that in the measurement, further indicating that such intensity in low **Q** regime cannot be solely from one-phonon coherent scattering and hinting the magnetic origin of these modes.

Temperature dependence of such spectral intensity at low **Q** is also consistent with the proposed magnetic origin. At elevated temperature, both static and dynamic correlations of magnetism are weakened by increased thermal fluctuations. Indeed, the magnetic INS cross section is directly related to the thermal average of spin correlations (Supplemental Information Eq. 3). Hence, it is natural to expect such anomalous intensities to weaken with increasing temperature if they have magnetic origin. On the other hand, INS spectral intensities from phonon lattice scattering are supposed to be stronger at high temperatures because, at low **Q** regime, the Debye-Waller factor contribution is trivial (see Supplementary Information Fig. S9) and the temperature-dependent lattice scattering cross section is only subject to the Bose-Einstein statistics (Supplemental Materials Eq. 1). As shown in Fig.2 a,b by the cropped slice, the measured INS spectrum in lower order BZs reveals dramatic weakening at elevated temperature. Such trend can also be observed in powder INS measurement (Fig.S3).

This trend is clearly illustrated by integrated mode intensities extracted from Lorentzian fitting of S(E) cuts at equivalent BZ boundaries. As shown in Fig.2 d1,e1, both TA and LA intensities decrease with increasing temperature at **Q** = (0,0,1). Again, this is unexpected because such descending trend cannot be explained by phonon scattering. More importantly, the spectral intensities of acoustic modes at (0,0,1) is stronger in AF phase and the intensities drastically decrease from 300 to 540 K through the phase transition temperature. This further suggests the anomaly is related to the magnetic order.

More strikingly, the pronounced intensity from TA mode, which is "forbidden" by the scattering geometry, is observed below 30 meV at 10 K (Fig.1 a, Fig.2 a, Fig.S1, Fig.S4, Fig.S5) in a broad range of **Q**. This feature is unexpected because the momentum transfer **Q** is perpendicular to the TA phonon eigenvectors **e** ($|\mathbf{Q} \cdot \mathbf{e}|^2 = 0$) for this orientation, making the lattice INS cross section zero for these modes (Supplementary Information Eq.1), except near Bragg points (Fig.1 b), where some intensity is expected due to finite integration width along other perpendicular **Q** directions. The observed intensity is not a result of the AFM-striction induced lattice distortion, which has been included in our scattering simulation. One plausible explanation is the formation of magnon polarons [24, 25, 28], which emerge from magnon-phonon hybridization and possess characteristics of both magnon and phonon. Because the magnon group velocity is much larger than that of phonon, the intersections of magnon and acoustic phonon dispersion in this system only exist in a small range of **Q** in the vicinity of the lattice BZ boundaries (magnetic BZ centers). Because the "forbidden" intensity is found not only around BZ boundaries but also elsewhere in reciprocal space without magnon-phonon crossings, the intensity cannot be solely from the magnon-phonon hybridization. Moreover, ignoring the small rhombohedral lattice distortion in AFM phase, the symmetry of all phonons at long wavelength limit has a irreducible representation of $\Gamma_{15}$ in Bouckaert-Smoluchowski-Wigner (BSW) notation [36]. Following the compatibility relations, the representation $\Gamma_{15}$ splits into $\Delta_1 \otimes \Delta_5$ along $[0,0,1]$ direction. In comparison, the magnon symmetry is of $\Delta_1' \otimes \Delta_2'$ [37], none of which are compatible with that of the phonon modes, as shown in Fig.1 c. As a result, magnons are not expected to hybridize with acoustic modes in NiO, and the anomalous spectral intensity from acoustic branches cannot be attributed to magnon-phonon hybridization.

Also surprising is that at higher order BZs, the temperature dependence of INS intensity from TA modes still shows a descending trend, suggesting its correlation to the magnetic order. As shown in Fig.2 d2-d4,e2-e4, at equivalent BZ boundaries of higher **Q** at $L > 2$, TA modes behave similarly to that in the first BZ and its intensity decreases with temperature, whereas LA intensity increases monotonically. This suggests that, at higher **Q**, the weakening of "forbidden" TA modes with temperature still shows their relation to the magnetic order. On the other hand, the temperature dependence of LA intensity at L = 3, 5, 7 agrees well with the phonon INS simulation, showing normal phonon behavior (Fig.2 e2-e4).

Although the appearance of phonon INS intensity at small **Q** is reminiscent of magneto-vibrational scattering (MVS), it cannot be modeled by MVS cross section. As part of magnetic INS, the MVS is elastic in spin system, inelastic in phonon system, and proposed based on the assumption of no correlation between lattice and spin [38]. While the MVS cross section has the same $\mathbf{Q}^2$ dependence as coherent one phonon scattering, it also contains a term related to the magnetic form factor $|F(\mathbf{Q})|^2$, giving weaker intensity at small **Q** (see Supplementary Information). A detailed comparison of phonon spectral intensities between the experiment and MVS models is shown in Fig.3 a,b. Clearly, the calculated lattice + MVS (see Supplementary Information) still fails for lower order BZs, suggesting the MVS model cannot satisfyingly explain the observations. On the other hand, the appearance of low-**Q** anomalous intensity cannot originate from neutrons scattered by phonon orbital magnetic moments [39-41], because they are only on the order of nuclear magneton [40], and the corresponding magnetic cross section will be 6 orders of magnitude smaller than that of typical magnetic INS by electronic dipoles. Therefore, anomalous intensity in the low-**Q** region may still originate from magnetic INS by electronic dipoles through a modified MVS process, in which lattice and spin are strongly correlated.

A magnetoelastic-correlated picture, in which the atomic displacements induced by phonon modulates the magnitude of magnetic moment (spin precession driven by phonon), explains the phonon INS intensity anomalies at small **Q**. Following the methodology discussed in Ref. [32], a modified MVS (mMVS) model, which contains an extra term related to the driving coefficient, was derived (see Supplementary Information). By fitting the driving coefficient $\xi_{LA}$ to the experimental data, the obtained calculated lattice + mMVS intensity can reproduce the experimental LA intensity at lower order BZs (Fig.3 a). This indicates that the observed anomaly can be attributed to the effect of spin precession driven by phonons and reveals strong dynamic correlations between magnetic moment and phonon induced lattice displacements. Meanwhile, one may expect such effect applies to TA phonons even though they are "forbidden" by the scattering geometry. This is because the mMVS model is non-zero under non-trivial driving coefficient even when $|\mathbf{Q} \cdot \mathbf{e}|^2 = 0$ [32]. Moreover, the driving coefficient of TA and LA modes can be similar in magnitude. This can be deduced from the non-collinear frozen phonon calculations (presented in Supplementary Information), which reveal the similar driving effects of TA and LA phonon induced atomic displacements to the magnitude of magnetic moment (Supplementary Information Fig.S6 a). This effect is analogous to the typical magnetoelastic coupling through dynamic modulation of exchange coupling strengths, which are of similar scales among various phonon branches by DFT calculation in NiO [42]. The driving coefficient for TA modes, $\xi_{TA}$, is obtained by fitting the experiment data, and its magnitude ($\xi_{TA} \approx 0.7\xi_{LA}$) reasonably agrees with the frozen phonon calculations (Supplementary Information Fig.S6 a). As shown in Fig.3 b, the calculated lattice + mMVS intensity successfully reproduces the TA intensity at small **Q**. Henceforth, the anomalously strong TA intensity at small **Q** originates from the effects of spin precession driven by phonons like LA.

However, at large **Q**, the appearance of the "forbidden" TA intensity cannot be modeled by mMVS. This is because the mMVS cross section is subject to $|F(\mathbf{Q})|^2$, and will approach zero at large **Q** ($L > 4$) (Fig.3 a,b). In fact, similar "forbidden" phonon modes have been observed in $Fe_{65}Ni_{35}$ [31, 32] by INS. Despite the success of mMVS model in explaining the "forbidden" TA modes in $Fe_{65}Ni_{35}$, the observed **Q**-dependence is completely different in NiO. While in $Fe_{65}Ni_{35}$ such mode shows a decrease in intensity at higher **Q**, our measurement presents an ascending trend, as can be seen in Fig.3 b. Clearly, the lattice + mMVS model still fails for TA modes at large $\mathbf{Q}$ (L > 4). In contrast, the experimental intensity from LA modes at L > 4 was successfully reproduced by the lattice scattering simulation, indicating that the LA intensity at larger $\mathbf{Q}$ is predominantly from lattice INS by phonons and is consistent with temperature dependent analysis above. The gigantic discrepancy between experimental intensity from TA modes and lattice scattering simulation indicates that the appearance of such "forbidden" TA modes is related to the spin-phonon coupling which is beyond the scope of mMVS model. Indeed, the effect of spin precession driven by phonon can only be reflected by magnetic neutron scattering at low $\mathbf{Q}$, thus it cannot explain those neutron scattering signatures of phonon eigenvector renormalizations shown at high $\mathbf{Q}$.

In the case of NiO, the "forbidden" TA intensity at L > 4 must predominantly result from lattice scattering instead of magnetic scattering and is suggestive of phonon eigenvector renormalization. The lattice INS origin is because magnetic INS intensity is always weaker at large $\mathbf{Q}$, following the magnetic form factor. From the experiment data shown in Fig.1 a, magnon spectral intensity decrease with the increase of $\mathbf{Q}$ and vanish at L > 4. It should be noted that the lattice counterpart of MVS (neutrons create or annihilate magnetic excitations via lattice scattering) can be safely ignored because the hyperfine coupling between the nuclear and electronic moments are weak [38]. The scattering intensity of this mode may have similar scattering origin as the "forbidden" intensity observed in iron chalcogenides [34]. For iron chalcogenides, the "forbidden" intensity vanishes under the spin-flip channel by spin polarized INS measurements, indicating it primarily originates from lattice INS by phonons. Therefore, the "forbidden" TA phonon INS intensity at high $\mathbf{Q}$ must predominantly originate from lattice INS by phonons. It is worthwhile mentioning that, such intensity cannot result from the instrument resolution or $\mathbf{Q}$ integration (Supplementary Information Fig.S7). If such INS intensity has a pure lattice origin, a renormalization of phonon eigenvector is necessary to explain the observed "forbidden" modes.

Renormalization of phonon eigenvector is usually associated with the change of phonon eigen-energy, which majorly comes from lattice thermal expansion, phonon-phonon, and spin-phonon couplings in magnetic insulators. To estimate the phonon energy renormalization from spin-phonon coupling, calculations based on quasi-harmonic approximation (QHA) are carried out to evaluate energy change contributed by lattice thermal expansion in the AFM spin configuration (see Methods). Comparing experimental acoustic phonon energy at BZ boundary with QHA calculations, the phonon energy of TA (LA) shows a 4%-5% (1%-2%) softening at 640 K (Fig.3 c,d). Moreover, the TA phonon energy dramatically decreases above $T_N$, showing a coincidence with the intensity change of the "forbidden" TA modes across $T_N$. This suggests both phonon energy and polarization are renormalized by the spin-phonon coupling.

The renormalization effect may originate from direct coupling between phonons and the local magnetization on ions. Although such renormalization process is related to the symmetry breaking because phonon eigenvectors follow the symmetry of the lattice, it's worthwhile emphasizing that, the static distortion induced by the AFM-striction cannot explain the observed "forbidden" TA modes, henceforth the symmetry breaking needs to be dynamic, as was pointed out in [34]. An early work suggested that the "forbidden" INS intensity in $Fe_{65}Ni_{35}$ [31, 32] may result from slow local orthorhombic distortions [43], which modulate the local magnetization of magnetic ions, then indirectly modify the dynamic matrix through magnetoelastic coupling, thereby cause the renormalization of eigenvectors. The modulation on local magnetization will result in an effective dynamical symmetry breaking of the phonon system and the corresponding "forbidden" intensity may originate from spin driven effects on lattice vibrations. In NiO, the orbital is not fully quenched (the ratio between orbital and spin moment $\frac{L}{S} = 0.34$) [44, 45], so the local magnetization includes both spin and orbit parts. In addition, previous INS studies reporting observations of "forbidden" phonons suggested that perturbations to orbital states can be critical in affecting phonon characteristics [33] and the renormalization effect may be related to the coupling between phonons and electron orbit degrees of freedom [34]. In the present case, the renormalization of phonon eigenvector can result from an effective dynamical symmetry breaking, which modify the dynamic matrix through magnetoelastic coupling, and is driven by the coupling between phonon and the spin and orbital states on the $Ni^{2+}$ ions.

Importantly, this can be reflected by the in-zone intensity of TA modes (Fig.3 b), which is found to be maximal at BZ boundaries ($L = 1, 3, 5, 7$). The TA modes at BZ boundary are non-propagating and shares the same spatial periodicity as the ground state magnetic order. Therefore, the renormalization effect is expected to be the most prominent at BZ boundaries because the coupling between phonons and sublattice magnetization is in phase. The abrupt transition from long-range spin order to short-range spin order across $T_N$ results in the weakening of spin-phonon coupling, and thus dramatically weakens the anomalous intensity in PM phase. However, the renormalization effect is not absent above $T_N$ because the short-range magnetic order still exists. Similar to the observed coherent scattering intensity of magnon at 540, 640K (Fig. S3 c3-c4), the "forbidden" intensities of TA modes exist above $T_N$.

Magnetic order induced anharmonicity of the phonon potential, magnon-phonon hybridization, electronic excitation-phonon couplings, and the presence of magnetic domains, are ruled out as possible origins of the phonon renormalization. Firstly, while the renormalization of eigenvectors can be closely related to the anharmonicity of the phonon potentials [46], frozen phonon calculations show that the phonon potentials of both TA and LA modes are quite harmonic regardless of their spin configuration (Supplementary Information Fig.S6 b). Secondly, renormalization of phonon eigenvectors have been found to be related to the interaction between phonons and magnetic excitations (magnons) in bulk YIG [47] and $YbB_{12}$ [33, 48]. In $YbB_{12}$, the anomalous temperature-dependent phonon INS intensity was successfully explained by the symmetry compatibility between phonons and magnons, the latter of which were assumed to be of the same symmetry as crystal field excitations. However, this is not the case for NiO, in which the symmetry of phonon and magnon are not compatible, and thereby prevents hybridizations of magnon and phonon modes, as mentioned previously. Finally, even though the symmetry of TA phonons ($\Delta_5$) and the first crystal field excitation state ($\Gamma'_{25}$) are compatible, the first crystal field excitation state (~1 eV) cannot be excited thermally in the studied temperature range, henceforth phonons are not expected to couple with the crystal field excitations. Further, the presence of magnetic domains should not affect the observed anomaly in the scattering intensity, as shown in Fig.S8. Due to the instrument energy resolution limitation, magnon gaps and their correlation to the observed anomalies cannot be characterized.

**Conclusion**

In summary, our measurement and simulations reveal the INS signature of mutual driving effects through strong spin-phonon coupling and acoustic phonon eigenvector renormalization in NiO for the first time. In particular, the measured anomalous INS intensity that follows the dispersion of acoustic phonon first weakens, then strengthens with increasing **Q**, suggesting a combination of magnetic and lattice scattering. The intensity at low **Q** is described by the mMVS model, unveiling the presence of spin precession driven by phonon. A spin-phonon driving coefficient determined by fitting the experiment data is used to quantify spin-phonon interaction strength. Additionally, the renormalization of phonon eigenvectors indicated by "forbidden" intensity at high **Q** is related to the magnetic order by its anomalous temperature dependence. Such renormalization may result from the coupling between phonon and local magnetization via spin driven effects on lattice vibration. Our study sheds light on the controlling of spin and lattice dynamics through spin-phonon couplings in antiferromagnetic spintronic materials. The mutual spin-phonon driving effects and the renormalization of phonon eigenvector may deserve investigation in other magnetic insulators, particularly those with strong spin-phonon coupling.

**Method**

**INS measurement.** Time-of-flight INS measurements were performed on single crystal NiO with the Wide Angular Range Chopper Spectrometer (ARCS) at the Spallation Neutron Source (SNS). Sample was placed on an Al holder and mounted in low-background electrical resistance vacuum furnace. Four-dimensional dynamic structure factors $S(\mathbf{Q}, E)$ were obtained at T = 10, 300, 540, and 640 K using incident energy of 150 meV, which covered multiple BZs and measured magnon and phonon simultaneously. Two extra measurements at 10, 300K were done with 5T magnetic field applied along [110] direction perpendicular to the scattering plane. Data reduction was done with MANTID [49]. The data was normalized by the proton current on target and corrected for detector efficiency. Since no detectable difference can be found in binning experimental data (10, 300K) with distorted rhombohedral or FCC lattice coordinates, the slight structure distortion in the AFM phase was neglected and the FCC crystal structure was used for data analysis. The data was sliced along high symmetry **Q**-directions in reciprocal space to produce two-dimensional energy-momentum views of dispersions.

Time-of-flight INS measurements were also performed on polycrystalline NiO. The sample was loaded in an Al sample can and mounted in a low-background electrical resistance vacuum furnace. Two-dimensional dynamic structure factors $S(|\mathbf{Q}|, E)$ were obtained at T = 50 and 640 K using incident energy of 50 meV and 150 meV. INS measurements on an empty Al can were performed at the same temperatures and the measured intensity, as the INS background induced by the sample holder, was subtracted from the polycrystalline data.

**INS data folding.** Data folding was used to increase counting statistics and remove the neutron scattering form factor in the dynamic structure factors $S(\mathbf{Q}, E)$. The data folding was done by summing up the $S(\mathbf{Q}, E)$ data from over 100 BZs into an irreducible wedge in the first Brillouin zone. Offsets of the *q* grid were corrected by fitting the measured Bragg diffractions. This folding technique has been used in the previous study [50] and proved to be reliable.

**First-principles calculations.** The ab initio density functional theory (DFT) calculations were performed with the VASP (Vienna Ab initio Simulation Package) [51, 52] on a plane-wave basis set, using the projector augmented wave (PAW) pseudopotentials [53, 54] with local spin density approximation (LSDA) exchange correlation functionals [55] and the Hubbard-U model [56]. U = 5 eV was chosen to obtain a best match with experimental phonon dispersion although it underestimated the electron band gap [57]. An energy cutoff of 550 eV was used for all calculations. LSDA+U ionic relaxation was done based on a primitive cell containing 2 nickel and 2 oxygen atoms with collinear antiferromagnetic (AFM) spin order. A Gamma-centered k-point grid of $13 \times 13 \times 13$ was used in LSDA+U ionic relaxation. The relaxed cell has a slight contraction along [1,1,1] direction. The calculated distortion angle of 0.15 °, which deviate from 60 ° in the FCC primitive cell, is larger than the experiment value in Ref [15]. The relaxed rhombohedral cell (space group 166) with a lattice constant of 4.95 Å and an angle of 33.66 ° (from an undistorted value of 33.55 °) was used in phonon dispersion calculations. The static dielectric tensor and Born effective charges were obtained to calculate non-analytical term in phonon calculations. The second order interatomic force constants were obtained from a $2 \times 2 \times 2$ supercell of 32 atoms with a Monkhorst-Pack k-point grid of $6 \times 6 \times 6$ using the density functional perturbation theory (DFPT). Phonon eigenvalues and eigenvectors were obtained by diagonalizing the dynamical matrix as implemented in the Phonopy [58]. The atomic mean squared displacements at various temperatures, and the projected phonon density of state were obtained based on the calculated phonon dispersion with a $q$-point sampling mesh of $30 \times 30 \times 30$.

Phonon dispersion calculations based on quasi-harmonic approximation (QHA) were carried out to estimate phonon energy change induced by lattice thermal expansion. The collinear AFM spin order was applied to all QHA calculations. The QHA calculations were based on measured thermal expansion at 10, 300, 540, 640 K. Specifically, the LSDA+U relaxed lattice constant was used as the value at 0 K, and the lattice constants at other temperatures were determined by thermal expansion measurements from Ref.[59]. The QHA calculations followed the same configurations discussed above.

To simulate modulations of the magnetic moment and the frozen phonon potential associated with TA and LA modes under different spin configurations, LSDA+U calculations were performed with atoms in the supercell displaced according to the eigenvectors of TA and LA modes at BZ boundary along [1,1,1]. Atomic displacements were set to be smaller than the calculated atomic root mean squared displacements at T = 500 K. Three types of spin configurations, standard collinear spin order (STD), non-collinear spin order with spin-orbit coupling (NCL), and non-spin-polarized (NSP) configurations, were considered. Under NCL spin configuration, the spin and orbit quantization axis were set to [1,1,-2] based on previous experiment results [60, 61]. For the original cell without atomic displacements, the obtained ratio between orbit and spin moment on the nickel ion was 0.1, smaller than the experiment value 0.34 [44]. Modulations of the on-site magnetic moment and frozen phonon potentials were obtained by varying displacement magnitudes under different spin configurations with a Gamma-centered k-point grid of $13 \times 13 \times 13$.

**Acknowledgments**

Q.S., Y.S., and C.L. are supported by the National Science Foundation under Grant No 1750786. This research used resources at the Spallation Neutron Source, a DOE Office of Science User Facility operated by the Oak Ridge National Laboratory.

**References**

[1] K. Uchida, H. Adachi, T. An, T. Ota, M. Toda, B. Hillebrands, S. Maekawa, E. Saitoh, Long-range spin Seebeck effect and acoustic spin pumping, Nat Mater, 10 (2011) 737-741.
[2] M. Weiler, H. Huebl, F.S. Goerg, F.D. Czeschka, R. Gross, S.T. Goennenwein, Spin pumping with coherent elastic waves, Phys Rev Lett, 108 (2012) 176601.
[3] A.V. Azovtsev, N.A. Pertsev, Magnetization dynamics and spin pumping induced by standing elastic waves, Physical Review B, 94 (2016) 184401.
[4] H. Hayashi, K. Ando, Spin Pumping Driven by Magnon Polarons, Phys Rev Lett, 121 (2018) 237202.
[5] J. Puebla, M. Xu, B. Rana, K. Yamamoto, S. Maekawa, Y. Otani, Acoustic ferromagnetic resonance and spin pumping induced by surface acoustic waves, Journal of Physics D: Applied Physics, 53 (2020) 264002.
[6] X. Zhang, G.E.W. Bauer, T. Yu, Unidirectional Pumping of Phonons by Magnetization Dynamics, Phys Rev Lett, 125 (2020) 077203.
[7] S.M. Rezende, D.S. Maior, O. Alves Santos, J. Holanda, Theory for phonon pumping by magnonic spin currents, Physical Review B, 103 (2021) 144430.
[8] T. Jungwirth, X. Marti, P. Wadley, J. Wunderlich, Antiferromagnetic spintronics, Nat Nanotechnol, 11 (2016) 231-241.

[9] G.E. Bauer, E. Saitoh, B.J. van Wees, Spin caloritronics, Nat Mater, 11 (2012) 391-399.
[10] S.R. Boona, R.C. Myers, J.P. Heremans, Spin caloritronics, Energy & Environmental Science, 7 (2014) 885-910.
[11] H. Wang, C. Du, P.C. Hammel, F. Yang, Antiferromagnonic spin transport from Y3Fe5O12 into NiO, Phys Rev Lett, 113 (2014) 097202.
[12] C. Hahn, G. de Loubens, V.V. Naletov, J. Ben Youssef, O. Klein, M. Viret, Conduction of spin currents through insulating antiferromagnetic oxides, EPL (Europhysics Letters), 108 (2014) 57005.
[13] H. Wang, C. Du, P.C. Hammel, F. Yang, Spin transport in antiferromagnetic insulators mediated by magnetic correlations, Physical Review B, 91 (2015) 220410.
[14] T. Shang, Q.F. Zhan, H.L. Yang, Z.H. Zuo, Y.L. Xie, L.P. Liu, S.L. Zhang, Y. Zhang, H.H. Li, B.M. Wang, Y.H. Wu, S. Zhang, R.-W. Li, Effect of NiO inserted layer on spin-Hall magnetoresistance in Pt/NiO/YIG heterostructures, Applied Physics Letters, 109 (2016) 032410.
[15] A.M. Balagurov, I.A. Bobrikov, S.V. Sumnikov, V.Y. Yushankhai, N. Mironova-Ulmane, Magnetostructural phase transitions in NiO and MnO: Neutron diffraction data, JETP Letters, 104 (2016) 88-93.
[16] H. Uchiyama, S. Tsutsui, A.Q.R. Baron, Effects of anisotropic charge on transverse optical phonons in NiO: Inelastic x-ray scattering spectroscopy study, Physical Review B, 81 (2010) 241103.
[17] Y. Wang, J.E. Saal, J.-J. Wang, A. Saengdeejing, S.-L. Shang, L.-Q. Chen, Z.-K. Liu, Broken symmetry, strong correlation, and splitting between longitudinal and transverse optical phonons of MnO and NiO from first principles, Physical Review B, 82 (2010) 081104.
[18] F.B. Lewis, N.H. Saunders, The thermal conductivity of NiO and CoO at the Neel temperature, Journal of Physics C: Solid State Physics, 6 (1973) 2525-2532.
[19] J. Oh, M.D. Le, H.H. Nahm, H. Sim, J. Jeong, T.G. Perring, H. Woo, K. Nakajima, S. Ohira-Kawamura, Z. Yamani, Y. Yoshida, H. Eisaki, S.W. Cheong, A.L. Chernyshev, J.G. Park, Spontaneous decays of magneto-elastic excitations in non-collinear antiferromagnet (Y,Lu)MnO3, Nat Commun, 7 (2016) 13146.
[20] K. Park, J. Oh, J.C. Leiner, J. Jeong, K.C. Rule, M.D. Le, J.-G. Park, Magnon-phonon coupling and two-magnon continuum in the two-dimensional triangular antiferromagnetCuCrO2, Physical Review B, 94 (2016) 104421.
[21] M.G. Cottam, D.J. Lockwood, Spin-phonon interaction in transition-metal difluoride antiferromagnets: Theory and experiment, Low Temperature Physics, 45 (2019) 78-91.
[22] E. Aytan, B. Debnath, F. Kargar, Y. Barlas, M.M. Lacerda, J.X. Li, R.K. Lake, J. Shi, A.A. Balandin, Spin-phonon coupling in antiferromagnetic nickel oxide, Applied Physics Letters, 111 (2017) 252402.
[23] D.J. Lockwood, M.G. Cottam, The spin‐phonon interaction in FeF2 and MnF2 studied by Raman spectroscopy, Journal of Applied Physics, 64 (1988) 5876-5878.
[24] C. Kittel, Interaction of Spin Waves and Ultrasonic Waves in Ferromagnetic Crystals, Physical Review, 110 (1958) 836-841.
[25] T. Kikkawa, K. Shen, B. Flebus, R.A. Duine, K.I. Uchida, Z. Qiu, G.E. Bauer, E. Saitoh, Magnon Polarons in the Spin Seebeck Effect, Phys Rev Lett, 117 (2016) 207203.
[26] H. Man, Z. Shi, G. Xu, Y. Xu, X. Chen, S. Sullivan, J. Zhou, K. Xia, J. Shi, P. Dai, Direct observation of magnon-phonon coupling in yttrium iron garnet, Physical Review B, 96 (2017) 100406.
[27] F. Godejohann, A.V. Scherbakov, S.M. Kukhtaruk, A.N. Poddubny, D.D. Yaremkevich, M. Wang, A. Nadzeyka, D.R. Yakovlev, A.W. Rushforth, A.V. Akimov, M. Bayer, Magnon polaron formed by selectively coupled coherent magnon and phonon modes of a surface patterned ferromagnet, Physical Review B, 102 (2020) 144438.

[28] J. Li, H.T. Simensen, D. Reitz, Q. Sun, W. Yuan, C. Li, Y. Tserkovnyak, A. Brataas, J. Shi, Observation of Magnon Polarons in a Uniaxial Antiferromagnetic Insulator, Phys Rev Lett, 125 (2020) 217201.
[29] Y. Li, C. Zhao, W. Zhang, A. Hoffmann, V. Novosad, Advances in coherent coupling between magnons and acoustic phonons, APL Materials, 9 (2021) 060902.
[30] J. Holanda, D.S. Maior, A. Azevedo, S.M. Rezende, Detecting the phonon spin in magnon–phonon conversion experiments, Nature Physics, 14 (2018) 500-506.
[31] P.J. Brown, I.K. Jassim, K.U. Neumann, K.R.A. Ziebeck, Neutron scattering from invar alloys, Physica B: Condensed Matter, 161 (1990) 9-16.
[32] P.J. Brown, B. Roessli, J.G. Smith, K.U. Neumann, K.R.A. Ziebeck, Determination of the wavevector and temperature dependence of the `forbidden' mode in Invar using inelastic neutron scattering, Journal of Physics: Condensed Matter, 8 (1996) 1527-1538.
[33] P.A. Alekseev, J.M. Mignot, K.S. Nemkovski, A.V. Rybina, V.N. Lazukov, A.S. Ivanov, F. Iga, T. Takabatake, Interplay of low-energy phonons and magnetic excitations in the Kondo insulator YbB12, J Phys Condens Matter, 24 (2012) 205601.
[34] D.M. Fobes, I.A. Zaliznyak, J.M. Tranquada, Z. Xu, G. Gu, X.-G. He, W. Ku, Y. Zhao, M. Matsuda, V.O. Garlea, B. Winn, Forbidden phonon: Dynamical signature of bond symmetry breaking in the iron chalcogenides, Physical Review B, 94 (2016) 121103.
[35] Z. Xu, J. Schneeloch, J. Wen, M. Matsuda, D. Pajerowski, B. Winn, Y. Zhao, C. Stock, P.M. Gehring, T. Ushiyama, Spin-wave induced phonon resonance in multiferroic BiFeO$_3$, arXiv preprint arXiv:1803.01041, DOI (2018).
[36] L.P. Bouckaert, R. Smoluchowski, E. Wigner, Theory of Brillouin Zones and Symmetry Properties of Wave Functions in Crystals, Physical Review, 50 (1936) 58-67.
[37] A.P. Cracknell, S.J. Joshua, The space group corepresentations of antiferromagnetic NiO, Mathematical Proceedings of the Cambridge Philosophical Society, 66 (2008) 493-504.
[38] A.T. Boothroyd, Principles of Neutron Scattering from Condensed Matter, Oxford University Press2020.
[39] S. Park, B.J. Yang, Phonon Angular Momentum Hall Effect, Nano Lett, 20 (2020) 7694-7699.
[40] D.M. Juraschek, N.A. Spaldin, Orbital magnetic moments of phonons, Physical Review Materials, 3 (2019) 064405.
[41] M. Hamada, E. Minamitani, M. Hirayama, S. Murakami, Phonon Angular Momentum Induced by the Temperature Gradient, Phys Rev Lett, 121 (2018) 175301.
[42] P. Maldonado, Y.O. Kvashnin, Microscopic theory of ultrafast out-of-equilibrium magnon-phonon dynamics in insulators, Physical Review B, 100 (2019) 014430.
[43] S. Lipinski, K.U. Neumann, K.R.A. Ziebeck, The forbidden phonon mode in Fe-Ni Invar, Journal of Physics: Condensed Matter, 6 (1994) 9773-9780.
[44] V. Fernandez, C. Vettier, F. de Bergevin, C. Giles, W. Neubeck, Observation of orbital moment in NiO, Physical Review B, 57 (1998) 7870-7876.
[45] S.K. Kwon, B.I. Min, Unquenched large orbital magnetic moment in NiO, Physical Review B, 62 (2000) 73-75.
[46] X. He, D. Bansal, B. Winn, S. Chi, L. Boatner, O. Delaire, Anharmonic Eigenvectors and Acoustic Phonon Disappearance in Quantum Paraelectric SrTiO_{3}, Phys Rev Lett, 124 (2020) 145901.
[47] H. Matthews, R.C. LeCraw, Acoustic Wave Rotation by Magnon-Phonon Interaction, Physical Review Letters, 8 (1962) 397-399.
[48] A.V. Rybina, P.A. Alekseev, J.M. Mignot, E.V. Nefeodova, K.S. Nemkovski, R.I. Bewley, N.Y. Shitsevalova, Y.B. Paderno, F. Iga, T. Takabatake, Lattice dynamics and magneto-elastic coupling in Kondo-insulator YbB12, Journal of Physics: Conference Series, 92 (2007) 012074.

[49] O. Arnold, J.C. Bilheux, J.M. Borreguero, A. Buts, S.I. Campbell, L. Chapon, M. Doucet, N. Draper, R. Ferraz Leal, M.A. Gigg, V.E. Lynch, A. Markvardsen, D.J. Mikkelson, R.L. Mikkelson, R. Miller, K. Palmen, P. Parker, G. Passos, T.G. Perring, P.F. Peterson, S. Ren, M.A. Reuter, A.T. Savici, J.W. Taylor, R.J. Taylor, R. Tolchenov, W. Zhou, J. Zikovsky, Mantid—Data analysis and visualization package for neutron scattering and μ SR experiments, Nuclear Instruments and Methods in Physics Research Section A: Accelerators, Spectrometers, Detectors and Associated Equipment, 764 (2014) 156-166.
[50] D.S. Kim, O. Hellman, J. Herriman, H.L. Smith, J.Y.Y. Lin, N. Shulumba, J.L. Niedziela, C.W. Li, D.L. Abernathy, B. Fultz, Nuclear quantum effect with pure anharmonicity and the anomalous thermal expansion of silicon, Proc Natl Acad Sci U S A, 115 (2018) 1992-1997.
[51] G. Kresse, J. Furthmüller, Efficient iterative schemes for ab initio total-energy calculations using a plane-wave basis set, Physical Review B, 54 (1996) 11169-11186.
[52] G. Kresse, J. Furthmüller, Efficiency of ab-initio total energy calculations for metals and semiconductors using a plane-wave basis set, Computational Materials Science, 6 (1996) 15-50.
[53] P.E. Blochl, Projector augmented-wave method, Phys Rev B Condens Matter, 50 (1994) 17953-17979.
[54] G. Kresse, D. Joubert, From ultrasoft pseudopotentials to the projector augmented-wave method, Physical Review B, 59 (1999) 1758-1775.
[55] J.P. Perdew, A. Zunger, Self-interaction correction to density-functional approximations for many-electron systems, Physical Review B, 23 (1981) 5048-5079.
[56] V.V. Anisimov, J. Zaanen, O.K. Andersen, Band theory and Mott insulators: Hubbard U instead of Stoner I, Phys Rev B Condens Matter, 44 (1991) 943-954.
[57] S.L. Dudarev, G.A. Botton, S.Y. Savrasov, C.J. Humphreys, A.P. Sutton, Electron-energy-loss spectra and the structural stability of nickel oxide: An LSDA+U study, Physical Review B, 57 (1998) 1505-1509.
[58] A. Togo, I. Tanaka, First principles phonon calculations in materials science, Scripta Materialia, 108 (2015) 1-5.
[59] S. P. Srivastava, R. C. Srivastava, I. D. Singh, S. D. Pandey, P. L. Gupta, Temperature Dependence of Thermal Expansion and Infrared Lattice Vibrational Mode of Nickel Oxide, Journal of the Physical Society of Japan, 43 (1977) 885-890.
[60] J. Baruchel, M. Schlenker, K. Kurosawa, S. Saito, Antiferromagnetic S-domains in NiO, Philosophical Magazine B, 43 (2006) 853-860.
[61] M.T. Hutchings, E.J. Samuelsen, Measurement of Spin-Wave Dispersion in NiO by Inelastic Neutron Scattering and Its Relation to Magnetic Properties, Physical Review B, 6 (1972) 3447-3461.

**Figures and Tables**

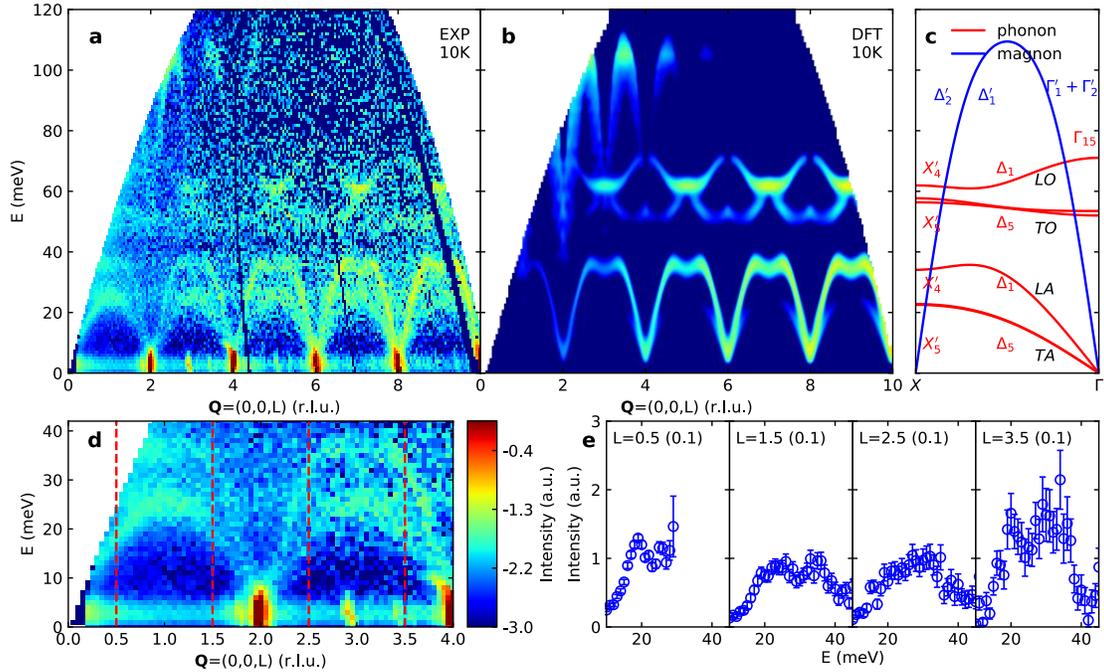

**Figure 1.** The measured and calculated dynamic structure factor, $S(\mathbf{Q}, E)$, of NiO at 10 K. (a, d) The dynamic structure factor of NiO measured by INS along the $[0, 0, 1]$ direction in the reciprocal space. The intensity is integrated over $\pm 0.1$ (r.l.u) along perpendicular axes and scaled by multiplying $E$. (b) Simulation of phonons and magnons with the same $\mathbf{Q}$ integration ranges and instrument resolution function. Both experimental data and theoretical calculations are plotted on logarithmic scale. (c) Calculated phonon and magnon dispersion along $\mathbf{\Gamma} - \mathbf{X}$ with BSW notation [36] for phonon and magnon [37]. Phonon branches that contribute to the scattering intensity are shown. (e) Constant $\mathbf{Q}$ cuts at various equivalent $\mathbf{Q}$ points, labelled by red dashed lines in (d).

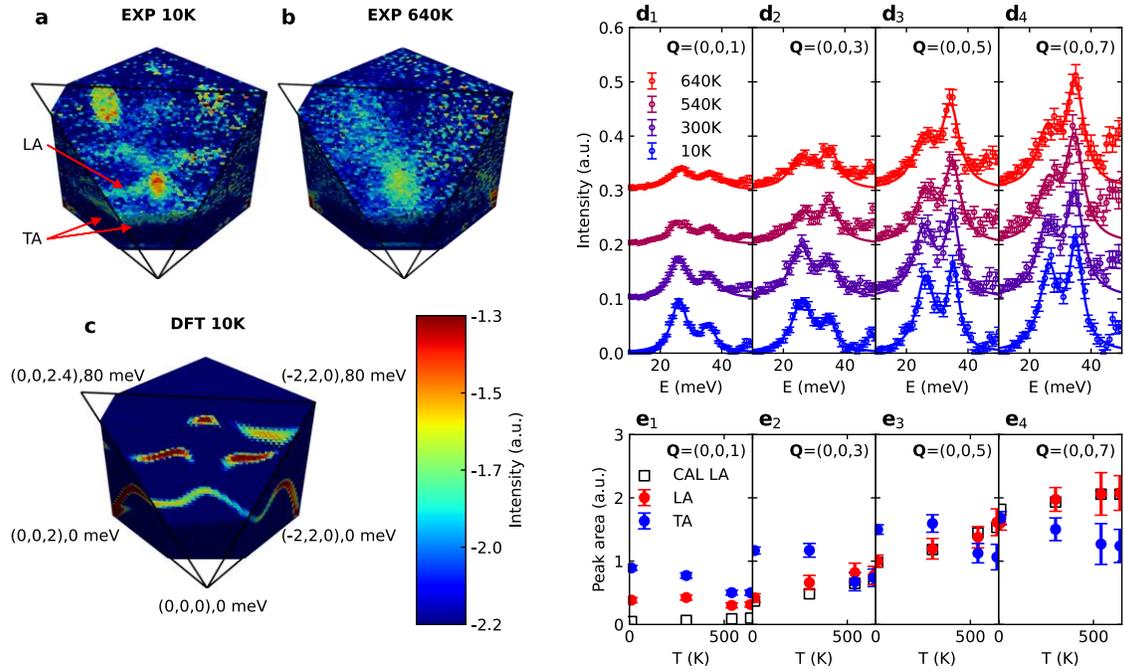

**Figure 2.** Anomalous temperature dependence of phonon INS intensity reveals strong coupling between acoustic phonon modes and spin. Volumetric view of the measured (a) (10 K), (b) (640 K), and calculated (c) (10 K) lattice scattering spectral intensity of coherent one phonon scattering in the (110) plane near $\mathbf{Q} = (0,0,0)$. Black lines indicate the limits of the cropped cross section. The intensity is in arbitrary unit. The spectral intensity of experiment data and calculation has been rescaled by multiplying E. (d1-d4) 1D spectral cuts at $L = 1, 3, 5, 7$ (r.l.u.) with $\mathbf{Q}$ integration ranges of $\mp 0.2$ on perpendicular directions. Symbols and colored curves represent experimental data and Lorentzian fits at 10, 300, 540, and 640 K. (e1-e4) Temperature dependence of mode intensity from Lorentzian fits of phonon modes at equivalent BZ boundaries along [0,0,1]. There is no magnetic Bragg peak along this direction, so that the spectral weights from magnons are negligible comparing to that of phonons in the experiment data below 40 meV and the peak areas represent the intensities of TA and LA, as is denoted by blue and red dots, respectively. Black squares represent the mode intensity from simulated dynamic structure factor with the same $\mathbf{Q}$ integration configurations for LA. Error bars indicate fitting errors.

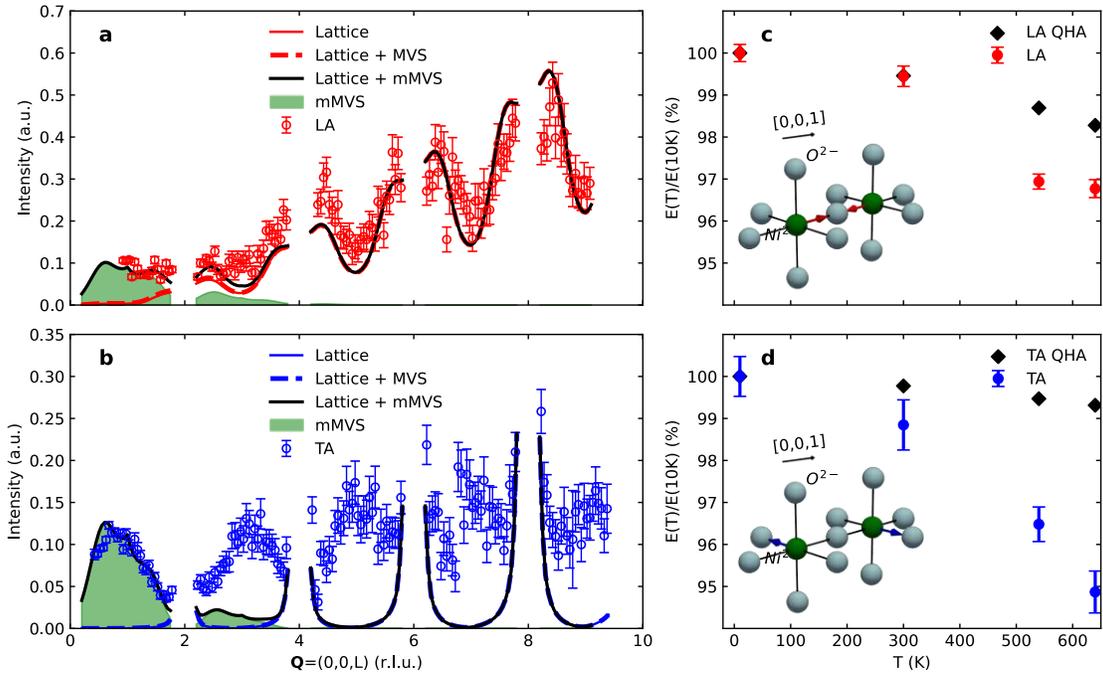

**Figure 3.** $Q$-dependence of TA and LA phonon INS intensity at 10 K and the temperature dependent TA and LA phonon energies at BZ boundary. (a,b) The $Q$-dependent spectral intensity comparisons between measurement and simulated $S(Q,E)$ of LA and TA modes along [0,0,1] are presented. The mode spectral intensities are obtained by subtracting a background and integrating a width of 13 meV following the calculated phonon dispersion. The red (Blue) circles represent experimental spectral intensities for LA (TA) modes. Data near Bragg points is masked. (c,d) The temperature-dependent acoustic phonon energy at BZ boundary plotted with quasi-harmonic approximation (QHA) calculations. Phonon energies are obtained from Gaussian fitting of the "folded" experiment data (see Supplementary Information). Red and blue dots denote relative LA and TA phonon energy E(T)/E(10K) at temperature T. The error bars denote fitting errors. Atomic motion corresponding to TA and LA modes at BZ boundary are sketched in the insets, where the red (blue) arrow indicate the phonon eigenvectors of LA (TA) and the black arrow indicates the phonon propagation direction in the real space.

16